\documentclass[amsmath,amssymb,aps,twocolumn,showpacs,prl,floatfix]{revtex4-1}
\usepackage{graphicx}
\usepackage{bm}
\usepackage{hyperref}
\usepackage{color}

\begin{document}

\title{Scaling quasi-stationary states in long range systems with dissipation}

\author{Michael Joyce$^1$, Jules Morand$^1$,  Fran\c{c}ois Sicard$^2$, and Pascal Viot$^3$}
\affiliation{$^1$Laboratoire de Physique  Nucl\'eaire et de Hautes \'Energies, UPMC IN2P3 CNRS  UMR 7585, 4, place Jussieu, 75252 Paris Cedex 05, France}
\affiliation{$^2$Laboratoire de Physique Th\'eorique, IRSAMC, CNRS UMR 5152 , Universit\'e Paul Sabatier, 31062 Toulouse, France}
\affiliation{$^3$Laboratoire de Physique
  Th\'eorique de la Mati\`ere Condens\'ee, UPMC, CNRS  UMR 7600, 4, place Jussieu, 75252 Paris Cedex 05, France}

\begin{abstract}
Hamiltonian systems with long-range interactions 
give rise to long lived out of equilibrium macroscopic states, so-called quasi-stationary states.
We show here that, in a suitably generalized form, this result
remains valid for many such systems in the presence of dissipation.
Using an appropriate mean-field kinetic description, we show that models with dissipation
due to a viscous damping or due to inelastic collisions admit ``scaling quasi-stationary states", i.e., states 
which are quasi-stationary in rescaled variables. A numerical study of one dimensional self-gravitating systems
confirms both the relevance of these solutions, and gives indications of their regime of
validity in line with theoretical predictions. We underline that the velocity distributions never 
show any tendency to evolve towards a Maxwell-Boltzmann form.
\end{abstract}
\bibliographystyle{apsrev4-1}
\maketitle
Physical systems characterized by  long range interactions (for reviews, see e.g. \cite{Campa2009,Bouchet2010}) 
are ubiquitous, encompassing systems as diverse as self-gravitating bodies in 
astrophysics, plasmas \cite{PhysRevLett.100.040604}, lasers \cite{Antoniazzietal_2006}, cold atoms \cite{Chalony2013} 
in the laboratory, and even biological systems \cite{sopiketal_2005}.
One  of the main results of recent years about such systems is that, quite generically, they relax, on 
times scales characterized by the mean force field, towards long-lived macroscopic states 
called quasi-stationary states (QSS) [e.g. galaxies in astrophysics \cite{Binney2008}, the red spot 
of Jupiter\cite{Bouchet_2002}, steady states of free electron laser\cite{Antoniazzietal_2006}].
These out-of-equilibrium states have a typical 
life-time diverging with particle number, while on shorter time scales they are described 
within the framework of the Vlasov equation. These results apply to  strictly
conservative  systems in a microcanonical framework, and the question inevitably 
arises of the robustness of such states beyond this idealized limit. Studies of
a paradigmatic toy model --- the Hamiltonian mean field (HMF)
model \cite{Antoni1995} --- have shown that, coupled to a canonical heat-bath \cite{Baldovin2009,Chavanis2011}, 
or when simple energy-conserving stochastic forces are introduced \cite{PhysRevLett.105.040602,Gupta2010},
such states relax rapidly towards thermal equilibrium. We report here theoretical and 
numerical results of the effect of introducing dissipative forces, with or without an intrinsic 
stochasticity.  Our main finding is that, for power law interactions, such systems
admit what we call ``scaling QSS",  i.e., solutions in which the phase space
distribution remains unchanged in rescaled variables as the system evolves.
Numerical study for a class of such models shows that these solutions
are often realized, and in the particular cases where deviations are observed,
the phase space density evolves with increasing correlation of velocity and 
position. This means in particular that these systems never shows any 
tendency, either in the scaling QSS or when there are deviations from them,
to evolve towards the space and (Maxwellian) velocity distributions of   
thermal equilibrium.

We consider particles interacting via a long-range  central 
power-law pair potential $V(r)=\frac{gm^2}{ n r^n}$ where 
$g$ is the coupling, $m$ the particle mass and $r$ the distance 
between the particles. The mean-field limit will be taken keeping  
the total energy $E$, total mass $M$ and system size $L$ fixed, 
with $N\rightarrow\infty$, and thus  $m \sim N^{-1}, g \sim N^0$.
For $n > 0$, the short distance cut-off should in general be regulated. 
We will not explicitly do so as we will treat the mean field dynamics 
which is in principle independent of the associated cut-off, at least 
down to  $n < d-1$ where $d$ is the dimension of space
\cite{PhysRevLett.105.210602}.

We consider in addition two different classes of dissipative forces: on the
one hand, a viscous damping force of the form $\vec{f}= - m \eta \| v \| ^{\alpha-1} \vec{v}$,
where $\alpha$ and $\eta$ are constants, which we will refer to as the viscous damping model (VDM); 
on the other hand, instantaneous inelastic, but momentum conserving, collisions, which we
will refer to as the inelastic collisional model (ICM). For the sake of simplicity, we restrict here to  one-dimensional models,
while the generalization to any dimension will be considered elsewhere. For two colliding 
particles $i$ and $j$  of incoming velocities $v_i$ and $v_j$, the post-collisional velocities are 
given by  $v^*_{i,j}=v_{i,j}\pm\frac{1+c }{2}(v_i-v_j)$,where  $c$ is the coefficient of restitution.
Amongst the many systems in this broad class, we note two particular ones. 
Firstly self-gravitating particles in an expanding universe are
described, in certain circumstances, in so-called comoving coordinates
and an appropriate time variable, by the case $\alpha=1$ corresponding
to a simple fluid damping $\vec{f}={-\eta m \vec{v}}$ (see e.g.
\cite{Joyce2011} and references therein). 
Secondly the case of gravity with inelastic collisions
corresponds to a self-gravitating granular gas. Piasecki and Martin \cite{Martin1996}  have 
obtained an exact solution of this model for specific regular initial conditions in
the totally inelastic limit, a situation different to that we will consider below.
Let us recall that, in the absence of gravity, i.e., for a simple granular gas starting from an 
homogeneous initial configuration, the kinetic energy of the 
system as well as the velocity distribution function have
been shown to obey scaling laws \cite{H83}, until the system 
reaches a collapse time where clusters appear\cite{Goldhirsch1993}.
An analogy between self-gravitating system and granular gases, was also 
considered for cluster formation by \cite{Fouxon2007,Shandarin1989,Aranson2006a}. 

In absence of dissipation, the time evolution is described, in the mean field limit, 
and thus on time scales short compared to that on which the full Hamiltonian evolution 
drives the system towards equilibrium, by the Vlasov equation
(see e.g. \cite{Chavanis2010, Bouchet2010})
\begin{align}\label{eq:bve1}
 \partial_t f(x,v,t)+v\partial_x f(x,v,t) +
\bar{a}(x,t) \partial_v f(x,v,t)= 0
\end{align}
where $\bar{a}(x)$ is the mean-field acceleration given by
$
\bar{a}(x,t)  =  g \int sgn(x-x')|x-x'|^{-(n+1)} f(x',v',t)dx'dv'
$
where $f(x,v,t)$ is the mass density in phase space.
When dissipative forces are present, the Vlasov equation 
is modified by the addition on the right hand-side of Eq.~(\ref{eq:bve1})  of a term denoted $J_d[f]$, an  operator 
accounting for the dissipation in the system \cite{Tuckerman1999}. For the case of a viscous damping force, this 
operator is \cite{arnold1989mathematical} 
\begin{equation}
 J_{d,1}=\eta \partial_v(  v^{\alpha} f(x,v,t)) \,,
\end{equation}
while for the case of inelastic collisions it can be expressed 
in terms of $f(x,v,t)$  (Stosszahl ansatz)\cite{PB03} as
\begin{align}
 J_{d,2}=\frac{N}{M}\!\int dv_1 |v-v_1|&\! \left[\frac{f(x,v^{**},t)f(x,v^{**}_1,t)}{c^2}
 \right.\nonumber\\
 &\left.
 -f(x,v,t)f(x,v_1,t)\right]
\label{Stosszahl ansatz}
\end{align}
where $v^{**}_i$ are the precollisional velocities which are given by
 $v^{**}_{i,j}=v_{i,j}\pm\frac{1+c^{-1}}{2}(v_j-v_i)$.  To obtain the
mean-field limit of Eq.~(\ref{Stosszahl ansatz}),  
 we rewrite the collision  
operator as a series expansion of $\frac{(1-c)}{2}$ \cite{PhysRevLett.107.138001}. 
After some calculation,  and taking the limit $N \rightarrow \infty$ at fixed 
$\gamma=\frac{(1-c)N}{2}$, we then obtain 
 $J_{d,2}=-\partial_v(  a_1(x,v,t) f(x,v,t)) \,,$
where $a_1(x,v,t)=\frac{\gamma}{M} \int du (u-v)|u-v|f(x,u,t)$ 
is the acceleration associated with the collisional force.
This scaling,  $(1 -c) \sim N^{-1}$, corresponds to the so-called quasi-elastic 
limit \cite{McNamara1993,Aumaitre2006}. As discussed below in detail,  in  this limit the ratio of the two essential time scales  of our system, the first
associated with the dissipation 
of the total energy 
and the second with the mean field dynamics ($\tau_{mf} \sim \frac{1}{\sqrt{g\rho_0}}$, where $\rho_0$ is the mass density), is independent of $N$.

We now seek scaling solutions to Eq.~(\ref{eq:bve1}), using the following ansatz:
\begin{equation}\label{eq:scaling}
 f(x,v,t)=\frac{M}{\bar{x}(t)\bar{v}(t)}F\left(\frac{x}{\bar{x}(t)},\frac{v}{\bar{v}(t)}\right)\,.
\end{equation}
Substituting this in Eq.~(\ref{eq:bve1}) gives
\begin{align}\label{eq:bvea1}
& -\frac{{\partial_t(\bar{x}(t)\bar{v}(t))}}{(\bar{x}(t)\bar{v}(t))^2} F(y,z)  
 -\frac{\partial_t\bar{x}(t)}{\bar{x}^2(t)\bar{v}(t)} y \partial_y F(y,z) -\frac{\partial_t\bar{v}(t)}{\bar{x}(t)\bar{v}^2(t)}\nonumber\\
&z \partial_z F(y,z) +\frac{z\partial_y F(y,z)}{\bar{x}^2(t)} + \frac{gM\bar{A}(y)}{\bar{v}^2(t)\bar{x}^{(n+2)}(t)}\partial_z F(y,z)=J_d
\end{align}
where $y$ and $z$ and the rescaled variables ($y=\frac{x}{\bar{x}(t)}$, $z=\frac{v}{\bar{v}(t)}$),  and  
$\bar{A} (y)=\int\!\mathrm{sgn}(y-y')|y-y'|^{-(n+1)}F (y',z)dy'dz$; 
Further we have for VDM $J_{d,1}=\frac{\eta\bar{v}^{\alpha-2}}{\bar{x}} \partial_z ( z^{\alpha} F(y,z)) $
and for ICM $ J_{d,2}=-\frac{\gamma}{\bar{x}^2(t)}\partial_z(\bar{A}_1 (y,z) F(y,z)) $
where $\bar{A}_1 (y,z)=\int (z'-z)|z-z'|F (y,z')dz'$.

These scaling solutions are admitted if it is possible 
to choose functions $\bar{x}(t)$ and $\bar{v}(t)$ so
that the time dependence of the coefficient of each
term is the same. Comparing, firstly, the last two terms 
on  the left-hand side
of Eq.~(\ref{eq:bvea1}), we infer the requirement
\begin{equation}
\bar{v}^2 (t) \bar{x}^{n}(t) = {\rm cste}\,.
\label{constantVR}
\end{equation}
This means that the virial ratio, defined
as 
$R=-\frac{2K}{nU}$
where $K$ and $U$ are the total 
kinetic and potential energy respectively, is 
constant. We now assume further
that  the system is in a QSS in the 
limit that the dissipation is absent.
With this assumption we have that
\begin{equation}
\label{eq:qss}
z\partial_y F(y,z) + \frac{gM}{\bar{v}^2 \bar{x}^{n}} \bar{A}(y) \partial_z F(y,z)=0\,.
\end{equation}
i.e., the last two terms on the right hand
side of Eq.~(\ref{eq:bvea1}) cancel. This also implies 
that the virial ratio $R$ is unity. Physically this means
we assume that all time dependence of the evolution
arises solely from the dissipation. This 
corresponds to an adiabatic limit  of  weak
dissipation in which the time scale on which
the dissipation causes macroscopic evolution
is arbitrarily long compared to the time scale 
associated with the mean-field dynamics. 
The scaling solution thus excludes 
all non-trivial time dependence due to the mean-field
dynamics beyond its effect in virializing the system,
and in particular therefore does not describe the 
phase of violent relaxation to virial equilibrium.
We will evaluate further below the validity of this
crucial approximation.

Using Eq.~(\ref{constantVR}) it is simple to infer that the sole additional 
requirement on the scaling solution is 
$\bar{v}(t)^{-1} \partial_t \bar{v}(t) =-A_0 \eta\bar{v}^{\beta}(t) $
with $\beta=\alpha-1$, for the VDM, and 
$\bar{v}(t)^{-1} \partial_t \bar{v}(t) =- A_1\gamma \bar{v}(t)\bar{x}(t)^{-1}
=-A_1\gamma \bar{v}^{\beta}$
with $\beta=(n+2)/n$ for the ICM, where $A_0$ and $A_1$ are dimensionless positive 
constants.
Integrating these equations,  we obtain
\begin{equation}
\bar{v}(t)=v_0\begin{cases}
            \left(1+\mathrm{sgn}(\beta)\frac{t}{t_c}\right)^{-\frac{1}{\beta}}\,& \beta \neq 0\\
             e^{-\frac{t}{t_c}}\,&\beta=0
           \end{cases}\label{eq:scaling_beta}
 \end{equation}
where $t_c$ is a characteristic time-scale. The solutions
for $\bar{x}(t)$ follow from Eq.~(\ref{constantVR}). 
The case $\beta=0$ corresponds to the VDM 
with $\alpha=1$, and the ICM with $n=-2$, the trivial case of a harmonic 
potential.

For a virialized state, the total energy $E=(1-\frac{2}{n})K$,
and so scales as $\bar{v}^2 (t)$. For attractive pair potentials with 
$n < 0$, which is the class of long-range potentials we are considering 
here (in $d=1$), the scaling solution therefore describes, for cases with
$\beta < 0$, a system which undergoes a collapse in the finite time $t_c$.
Otherwise, the system undergoes a monotonic 
contraction characterized by the
same time, but never collapses.

We now return to the essential approximation
Eq.~(\ref{eq:qss}) which we have made in deriving
the scaling solution. This corresponds 
to assuming that $\tau_{diss} \gg \tau_{mf}$,  
where  $\tau_{diss}$ and $\tau_{mf}$ are
the characteristic times for, respectively,  the dissipation 
of the system energy $E$ and the mean-field (Vlasov) 
dynamics. For the VDM we have 
that 
\begin{equation}
\frac{dE}{dt} = - \eta \langle |v|^{\alpha+1} \rangle_{t}
\end{equation}
where $\langle X(x,v) \rangle_{t}=\int dxdv X(x,v) f(x,v, t)$.
Substituting the scaling solution, in which $E\propto \bar{v}^2(t)$,
in this equation,
we can then infer that 
\begin{equation}
 t_c \equiv  \frac{2E_0}{\eta \langle |v|^{\beta+2} \rangle_0}= \frac{1}{\eta}\left(1-\frac{2}{n}\right) \frac{\langle v^{2} \rangle_0}{\langle |v|^{\beta+2} \rangle_0}
\label{tc}
\end{equation}
For the ICM, a similar relation can be written, the only
difference being that $\eta$ is replaced by $\gamma I_0$
where $I_0$ is a dimensionless integral. For both cases, 
$t_c$ diverges as the inverse of the strength 
of the dissipation. $t_c$ represents the time scale for dissipation starting 
from the (arbitrary) time $t=0$. In the scaling solution, the
characteristic time for dissipation of energy starting from
an arbitrary time $t$ thus scales as 
$\tau_{diss} (t)  \propto \bar{v}^{-\beta} (t)$.
For a typical system size  $\bar{x}(t)$,
the mean field acceleration scales as $\bar{x}^{-(n+1)}$, 
and   $\tau_{mf}$  (mean time for a
particle to cross the system)   as $\bar{x}^{\frac{n+2}{2}}$.
It follows that
$
\tau_{diss}/\tau_{mf}  \propto \bar{v}^{[-\beta+\frac{n+2}{n}]} 
$
and therefore if $\beta > \beta_c=\frac{n+2}{n}$ 
the ratio of these timescales increases as a function of time. In other
words, if $\beta > \beta_c$, the scaling solution
drives the system to a regime in which the approximation underlying 
it becomes arbitrarily well satisfied. In
this case, we then expect that the scaling solution may be an 
attractor for the system's behavior, while for $\beta < \beta_c$, 
the opposite is the case and the scaling solution is at most 
expected to represent a transient behavior. The case 
$\beta=\beta_c$, which corresponds precisely to the ICM,  is the marginal one. In this case, 
the ratio $\tau_{diss}/\tau_{mf}$ remains constant in the
scaling solution, and one would expect it to be a transient 
which persists on a time-scale dependent on 
this ratio.

\begin{figure}[t]
\includegraphics[scale=0.15]{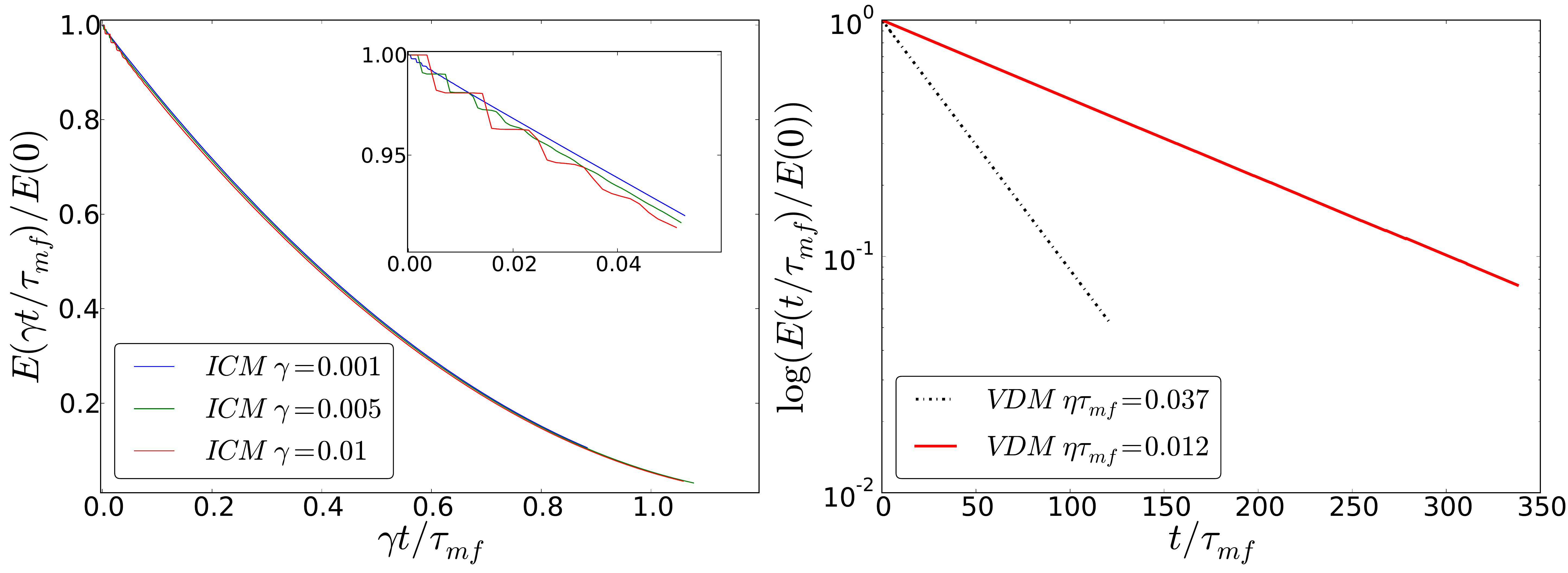} \\
 \caption{Left: For the ICM, reduced total energy $E(t)/E(0)$ versus $\gamma t/\tau_{mf}$, for 
the indicated values of $\gamma$ and $R_0=0.01$. Inset: zoom on the
shorter time evolution.
Right: for the VDM, semi-log plot of $E(t)/E(0)$  versus $ t/\tau_{mf}$ with the indicated values of 
$\eta$, and  $R_0=0.01$. }
\label{fig:decaygamma}
 \end{figure} 

To explore the validity of this analysis, we have performed a 
numerical study of two $1D$ self-gravitating systems, 
i.e. the case $n=-1$ corresponding to the pair potential
$\phi(x)=gm|x|$ derived from the 1D Poisson equation 
$\partial_x^2\phi(x)=2gm\delta(x)$, one with the 
dissipation of the ICM, and the other that of the VDM for 
the case $\alpha=1$. In the absence
of dissipation the equations of motion may  be
integrated exactly between particle collisions (equivalent to
crossings), and the system can be evolved using an 
event-driven algorithm\cite{Noullez2003,Yawn2003,Joyce2011} 
which determines the collision times exactly (up to round-off errors).
For the ICM, inelastic collisions are  implemented in an existing code 
with the appropriate post-collisional velocities (including the coefficient 
of restitution $c$).
For the VDM, an event-driven 
algorithm is also implemented as the collision time between particles 
in this case can also be computed exactly by finding the roots of a 
quintic equation (see \cite{Joyce2011} and references therein). 
We use ``rectangular water-bag'' initial conditions, i.e., velocities
and positions are chosen randomly and uniformly  in phase space 
in $[-v_0,v_0]\times [-L_0/2,L_0/2]$.
These are fully characterized by the initial 
virial ratio $R_0$. 
We have performed simulations of different sizes  ($N=256...4096$), and no noticeable finite effects have been 
observed.  All quantities have been averaged
over $100$ independent realizations for the ICM,
and $50$ for the VDM (which is less noisy).  
For both models, the simulation is interrupted   
when the difference between two possible collision times 
becomes smaller than the accuracy of the computer.

We define the mean-field time as $\tau_{mf}=2\sqrt{\frac{L_0}{gN}}$, 
and recall the behaviour of this system in absence of dissipation
from initial conditions of this kind (as detailed, e.g. in \cite{Joyce2010}):
it evolves on a time-scale of order $10 - 100$ $\tau_{mf}$
towards a QSS, in which the virial ratio $R$ is unity. Monitoring
$R$ in the present case  (with dissipation) we find essentially 
identical behaviour, but, as expected, a very different behaviour 
for the energy. 
The left panel of Fig.~\ref{fig:decaygamma} shows, for the ICM, the 
normalized energy  as a function of the dimensionless time $\gamma t/\tau_{mf}$, 
for an initial virial  ratio  $R_0=0.01$ and the different given values of $\gamma$; 
in the right panel the same quantity is plotted versus $ t/\tau_{mf}$,  for
 the VDM  with $\eta \tau_{mf}  = 0.012, 0.037$.
We observe excellent agreement with the scaling solutions: for the ICM, the 
energy decay is fitted by $(1-\frac{t}{t_c})^\delta$ 
with $\delta=2.00\pm0.01$;
for the VDM, the energy decay is fitted by 
$E(t_{s})/E(0)=\exp(-\frac{2}{3}\lambda\frac{t}{\tau_{mf}})$
with $\lambda= \eta \tau_{mf} $,
as predicted by the
scaling solution (Eqs. (\ref{eq:scaling_beta})
and (\ref{tc})), to $10^{-4}$.
The inset of Fig.~\ref{fig:decaygamma} shows small deviations from the scaling behaviour
at short times, associated with the virial oscillations during the 
initial violent relaxation.
\begin{figure}[th]
\includegraphics[scale=0.15]{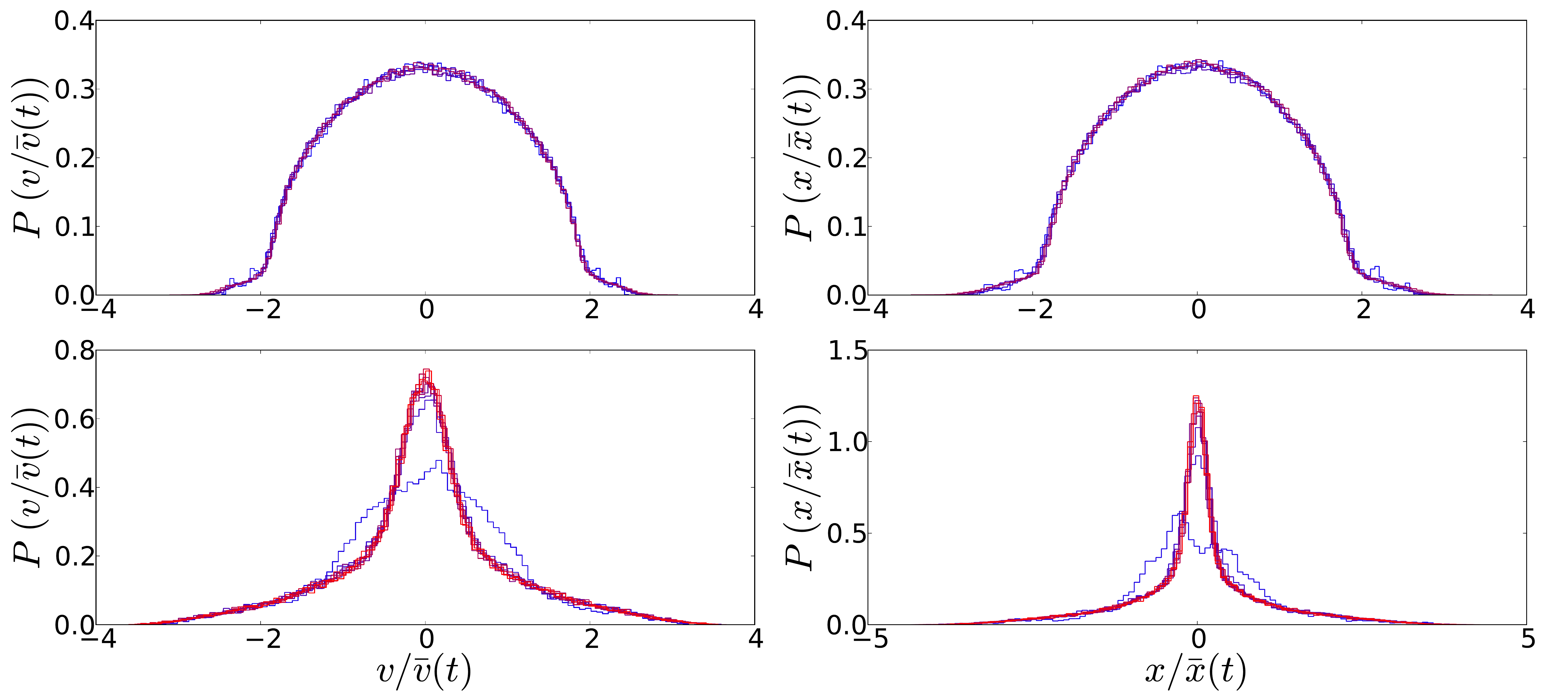}\\
 \caption{Velocity (left panels) and position (right panels) distributions as
a function of rescaled variables,  for the ICM with $\gamma=0.01$ at different times
$t/\tau_{mf}=10,31,53,74,95$ and  $R_0=1$ (upper panels) and  
for the VDM  with $\eta\tau_{mf}=0.037$ and $R_0=0.01$ (lower panels)  at
$t/\tau_{mf}=10,20,...,100$.}
\label{fig:rescaledvel}
\end{figure}

\begin{figure}[t]
\includegraphics[scale=0.15]{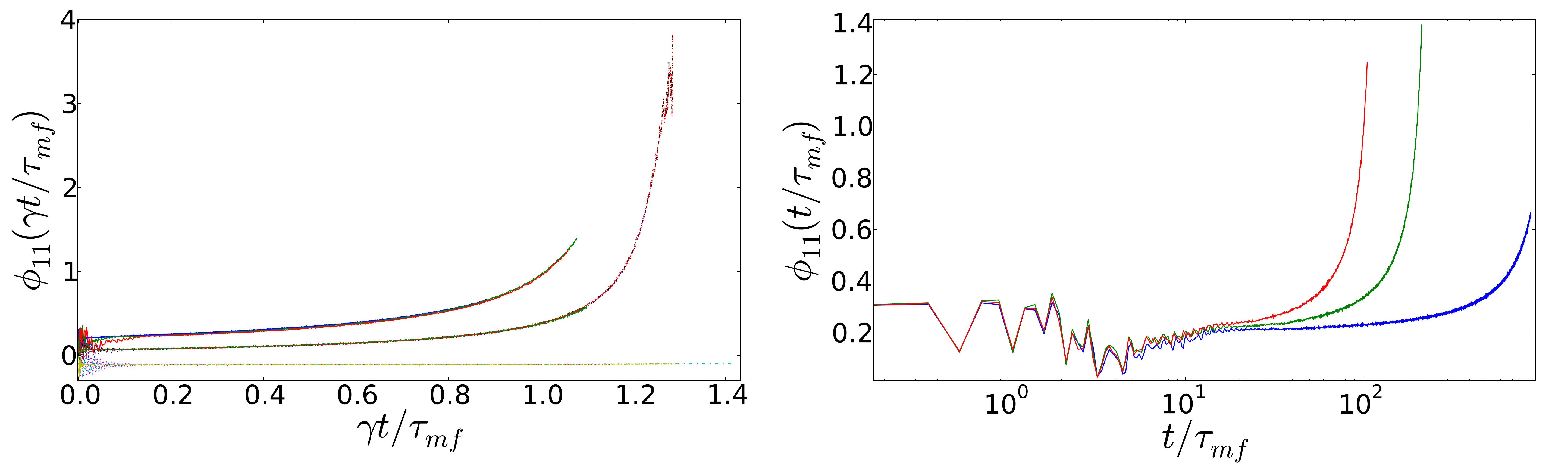} \\
 \caption{Left: $\phi_{11}$ versus $\gamma t/\tau_{mf}$ for $\gamma= 0.01, 0.005, 0.001$ and $R_0=0.01$ (3  upper curves), 
 $R_0=0.1$ (3 middle curves) and $R_0=1$ (3 lower curves).
Right: $\phi_{11}$ versus $t/\tau_{mf}$ in a semi-log plot for $R_0=0.01$ and $\gamma=0.001$ (red), $0.005$ (green), $0.01$ (blue)}
\label{fig:phi11gamma}
 \end{figure} 
 
The velocity and position distributions
versus appropriate rescaled variables are shown  in Fig.~\ref{fig:rescaledvel} at 
different times, for  the ICM with  $R_0=1$, and for the VDM with  $R_0=0.01$.
The superposition of the curves illustrates the accurate 
description of the kinetics by scaling QSS
(in the lower panels, the blue curves correspond to $t/\tau_{mf}=10$,
in the phase of the violent relaxation). Moreover, one observes 
significant deviations from a Gaussian shape of the distribution
corresponding to  the existence  of a  ``core-halo" structure in the
QSS \cite{Teles_2011,PhysRevE.84.011139}.

The ICM is a marginal case for the validity of the approximation in which
we obtained the scaling QSS, while for the VDM with 
$\alpha=1$ and $n=-1$ (i.e. $\beta_c=-1$), we expect it to become
exact asymptotically. To quantify deviations from the scaling QSS 
it is convenient to monitor the dimensionless quantity \cite{Joyce2010}
\begin{equation}
\phi_{11}=\frac{\left\langle |x||v|\right\rangle }{\left\langle |x|\right\rangle 
\left\langle |v|\right\rangle }-1=\frac{\int|x||v|fdxdv}{\int|x|fdxdv\int|v|fdxdv}-1 \,,\label{eq:phi11}
\end{equation}
which provides a measure of the correlation
between the spatial and velocity variables.
For conservative self-gravitating systems, $\phi_{11}$ can be interpreted as an order parameter for the QSS, which goes to $0$ as the system 
goes to thermal equilibrium\cite{Joyce2010}. Inserting Eq.~(\ref{eq:scaling}), 
in Eq.~(\ref{eq:phi11}), we see that $\phi_{11}$ is constant in time also in
the scaling QSS: the 
evolution of the system is through a sequence of QSS with identical correlations.

Figure \ref{fig:phi11gamma}, right panel shows  the evolution of $\phi_{11}$  
for the ICM starting from $R_0=0.01$, $R_0=0.1$ and $R_0=1$,
and for the three different values of $\gamma$. We have used here this rescaled time  $\gamma t/\tau_{mf}$
because we observe that it gives a good collapse of all curves;
on the left panel, the result for the first case ($R_0=0.01$) is shown without this rescaling.
For the VDM  (not shown here)  $\phi_{11}$  
remains constant (after the initial violent relaxation). 
We observe that while the case $R_0=1$, $\phi_{11}$  is almost constant, 
visible deviations are evident for the two other cases, with
evolution away from the scaling setting in fastest for the case $R_0=0.01$.
Similarly,  the space and velocity distributions deviate
progressively  from the scaling solutions for $R_0=0.1,0.01$ (not shown  here).
In both cases, the system energy decreases as predicted by the scaling solution. 
Moreover, this evolution 
appears to depend only on the QSS attained (which is different for each
$R_0$), and  on $\gamma$ through the rescaled variable. 
Thus, as the total energy  goes to $0$, inelastic collisions drive the system through 
a given family of  ever more correlated
QSS. 
It implies that the system never shows any tendency  to drive the system towards a 
Maxwell-Boltzmann distribution of velocities (nor towards the spatial distribution 
of the thermal equilibrium of the model), despite the effective stochasticity of the inelastic collisions. 
This  contrasts to what is observed in a stochastically perturbed 
HMF model in \cite{PhysRevLett.105.040602,Gupta2010}. This tendency towards 
more correlated states can be interpreted as follows: for $R_0 \ll 1$, the violent relaxation drives the system 
to a core-halo structure (see \cite{PhysRevLett.100.040604,Joyce2010,Teles_2011}), 
whereas for $R_0 \simeq 1$, the QSS is rather homogeneous in phase space. For the ICM,
the (kinetic) temperature of the core decreases more rapidly than that of the halo;
in the VDM, on the other hand, the systems cools down uniformly. Further it may
be, as observed in three dimensional gravitating systems \cite{Ispolatov2004}, that 
inefficiency of energy exchange between the core and halo impedes relaxation
towards thermal equilibrium.

The authors thank C. Rulquin for useful remarks.


%

\end{document}